\documentclass[10pt, preprint]{aastex}

\usepackage{natbib}
\usepackage{float}
\usepackage{graphicx}
\usepackage{subfig}
\usepackage{amsmath}
\usepackage[toc,page]{appendix}
\usepackage[utf8]{inputenc}
\usepackage{hyperref}

\usepackage{booktabs}
\renewcommand{\arraystretch}{1}

\title{Spectral Kurtosis Based RFI Mitigation for CHIME}

\author{
Jacob Taylor\altaffilmark{1}\altaffilmark{2}, 
Nolan Denman\altaffilmark{1}\altaffilmark{2},
Kevin Bandura\altaffilmark{3}\altaffilmark{4},
Philippe Berger\altaffilmark{5}\altaffilmark{6}, \\
Kiyoshi Masui\altaffilmark{7},
Andre Renard\altaffilmark{1}, 
Ian Tretyakov\altaffilmark{1}\altaffilmark{5}, 
Keith Vanderlinde\altaffilmark{1}\altaffilmark{2}
	}
\altaffiltext{1}{Dunlap Institute for Astronomy and Astrophysics, University of Toronto}
\altaffiltext{2}{Department of Astronomy and Astrophysics, University of Toronto}
\altaffiltext{3}{Department of Computer Science and Electrical Engineering, West Virginia University}
\altaffiltext{4}{Center for Gravitational Waves and Cosmology, West Virginia University}
\altaffiltext{5}{Department of Physics, University of Toronto}
\altaffiltext{6}{Canadian Institute for Theoretical Astrophysics, University of Toronto}
\altaffiltext{7}{Department of Physics and Kavli Institute for Astrophysics and Space Research, Massachusetts Institute of Technology}

\begin{document}

\begin{abstract}
We present the implementation of a spectral kurtosis based Radio-Frequency Interference detection system on the CHIME instrument and its reduced-scale pathfinder. Our implementation extends single-receiver formulations to the case of a compact array, combining samples from multiple receivers to improve the confidence with which RFI is detected. Through comparison between on-sky data and simulations, we show that the statistical properties of the canonical spectral kurtosis estimator are functionally unchanged by cross-array integration. Moreover, by comparison of simultaneous data from CHIME and the Pathfinder, we evaluate our implementation's capacity for interference discrimination for compact arrays of various size. We conclude that a spectral kurtosis based implementation provides a scalable, high cadence RFI discriminator for compact multi-receiver arrays.
\end{abstract}

\maketitle

\section{Introduction}
The presence of non-astronomical signals in the electromagnetic spectrum 
(`Radio-Frequency Interference' or RFI)
poses a significant hazard to 
successful radio observations.
Sources of such interference include lightning and electrical discharges, 
inadvertent emissions from terrestrial electronics, 
radio-frequency telecommunication signals, 
and communications with satellites and aircraft
\citep{1991ASPC...17..213G, 2013MNRAS.435..584O, 2016arXiv161004696S}.

The Canadian Hydrogen Intensity Mapping Experiment 
(CHIME\footnote{\url{http://www.chime-experiment.ca}})
is a newly-constructed radio interferometer
with 1024 dual-polarization receivers 
continuously observing a 400-800\,MHz band.
A fully independent reduced-scale prototype, 
the CHIME Pathfinder (henceforth `the Pathfinder')
possesses 128 dual-polarization receivers 
and an identical observing band \citep{2014SPIE.9145E..22B,2014SPIE.9145E..4VN}.
CHIME and the Pathfinder
each follow an `FX' correlator architecture,
with an FPGA-based Fourier-transform stage
\citep{2016JAI.....541005B}
followed by a GPU-based outer-product `X-engine'
\citep{2015arXiv150306203K,2015arXiv150306189R,2015arXiv150306202D}.

The considerable capabilities of the CHIME correlator present an opportunity for powerful, on-the-fly RFI mitigation.
CHIME employs a `software' correlator X-engine, and its easily-reconfigurable, general-purpose hardware permits the introduction of additional data processing with minimal development effort.
This includes extending the real-time processing system to include the detection and excision of RFI during the correlation process.
The architecture of the CHIME X-engine
is such that each processing `node' contains data from only a few frequencies, 
but over the entire array;
this constrains the selection of RFI-excision algorithms which may be applied in the GPUs.
Our choice of a statistical excision algorithm, based on spectral kurtosis, was informed by both robust performance and modest computational cost.
Additionally, as detailed below, the compact layout of CHIME permits multi-receiver RFI detection, lowering the effective threshold for RFI detections while retaining high time resolution.

We summarize the theoretical foundations of the spectral kurtosis estimator in \S\ref{sec:theory} and present a multi-receiver formulation of the estimator in \S\ref{ssec:array}. The details of its implementation on CHIME and the Pathfinder are included in \S\ref{ssec:imp} and the results obtained follow in \S\ref{ssec:results}. 

\section{The Spectral Kurtosis Estimator}
\label{sec:sk-gen}
 
Tests of the normalized higher moments of a distribution provide a powerful metric for the presence of RFI. Given the expected Gaussianity of the electric fields produced by natural sources,
and the extreme non-Gaussianity of observed RFI sources, deviations in the statistical properties of radio measurements signal the presence of substantial artificial contamination.

The use of a kurtosis-based estimator as an indicator of significant non-Gaussianity, and therefore contamination, has a long history: \citet{1172264, dwyer-oe, 681555, 2003-vrabie-sk} and \citet{2007PASP..119..805N} explore a spectral-domain kurtosis measure, with further development and radio-astronomical application in \citet{2010PASP..122..595N, 2010MNRAS.406L..60N, 2010PASP..122..560G}, and \citet{2016MNRAS.458.2530N}.
Spectral-domain kurtosis measurement provides a robust and overall-scaling-invariant probe of the data distribution, with substantial potential for discrimination.
 
\subsection{Spectral Kurtosis}\label{sec:theory}
For a set of $M$ independent complex values $x_j$, representing the post-Fourier-transform timestream of a single frequency channel as received by the X-engine, we construct an unbiased spectral kurtosis estimator $\widehat{SK}$ after \citet{2010PASP..122..595N}:

\begin{equation}
	S_1 = \sum_j^M |x_j|^2 \hspace*{20pt}
	S_2 = \sum_j^M |x_j|^4 
\end{equation}
\begin{equation}
	\widehat{SK} = \left(\frac{M+1}{M-1}\right)\left(M\frac{S_2}{(S_1)^2} - 1\right)
	\label{eq:sk}
\end{equation}

For $x_j$ drawn from a circularly-symmetric complex Gaussian, the $|x_j|^2$ will be $\chi^2$-distributed with two degrees of freedom.
Hence, as shown by \citet{2010PASP..122..595N}, to first order the expected mean, variance, skewness, and kurtosis of $\widehat{SK}$ are:
\begin{gather}
 \textrm{mean}(\widehat{SK}) = 1\\
 \textrm{variance}(\widehat{SK}) \approx \frac{4}{M} + \mathcal{O}(\frac{1}{M^2})\\
 \textrm{skew}(\widehat{SK}) \approx \frac{10}{\sqrt{M}} + \mathcal{O}(\frac{1}{M^\frac{3}{2}})\\
 \textrm{kurtosis}(\widehat{SK}) \approx \frac{246}{M} + \mathcal{O}(\frac{1}{M^2})
\end{gather}

Notably, the expected mean is invariant and the variance, skewness and kurtosis depend solely on $M$.

\subsection{Spectral Kurtosis in a Compact Array Context}
\label{ssec:array}
For an interferometer in which the receivers are separated by a maximum distance $L$, and in which the data from each receiver is used to compute a spectral kurtosis estimate over a period $\Delta t$,
any transient RFI signal will appear `simultaneous' across the array in the case that $\frac{L}{c} \ll \Delta t$.
This is true of CHIME's compact configuration: the receivers are separated by at most $\approx$\,100\,m, so any RFI signal will arrive at all the receivers within $\approx$\,0.3$\,{\mu}$s, 
far shorter than the timescales at which spectral kurtosis estimates are produced in our implementation.
For an array which is compact relative to the distance to sources of RFI,
receivers will see the interference at similar levels, 
permitting their combination to enhance RFI detection.

The discrimination potential of a spectral kurtosis based estimator is limited by its variance, which depends solely on the total number of samples present in the input data, $M$. Therefore, increasing $M$ corresponds to increasing the discrimination potential of the estimator. For a single receiver with a fixed sampling rate, improving the discrimination potential of the estimator necessarily reduces the cadence at which estimates are produced. Alternatively, combining measurements from multiple receivers can increase the size of input dataset (reducing the inherent variance in the estimator) without affecting the cadence at which kurtosis measurements are generated. If the receivers provide uncorrelated samples (see Appendix~\ref{appendix:AppendixA} for limitations on this) combining data from multiple receivers will increase the effective integration length, reducing the inherent variance in the estimator when calculated at a given cadence

\subsubsection{Multi-Receiver Formulation of the Spectral Kurtosis Estimator} \label{sec:MultiFormulation}

For a compact array of $N$ independent receivers, 
in which spectral kurtosis estimates are desired every $n$ time-samples,
one possible method of combining the information from the different antennas
would be to simply average the individual spectral kurtosis estimates after they are produced.
We present here an alternative formulation
which produces mathematically equivalent results
by combining the intermediate cumulants from different receivers into a single overall estimator.

We begin by considering the spectral kurtosis estimate produced by the $i^{th}$ receiver:

\begin{equation} \label{singlesums}
	(S_1)_i = \sum_j^n |x_j|^2 \hspace*{20pt}
	(S_2)_i = \sum_j^n |x_j|^4 
\end{equation}

\begin{equation}\label{originalSK}
    \widehat{SK}_i = \frac{n+1}{n-1} \left ( \frac{n(S_2)_i}{((S_1)_i)^2} - 1 \right)
\end{equation}

After normalizing each receiver's signal by its mean power $\mu_i$, we sum across the array to produce the normalized cumulants $ \overline{S_1}$ and $\overline{S_2}$:

\begin{equation}\label{discreteMean}
    \mu_i= \frac{(S_1)_i}{n}
\end{equation}

\begin{equation}\label{normedSum1}
    \overline{S_1} = \sum^N_i \frac{1}{\mu_i}(S_1)_i = nN
\end{equation}

\begin{equation}\label{normedSum2}
    \overline{S_2} = \sum^N_i \frac{1}{(\mu_i)^2}(S_2)_i = \sum^N_i \frac{n^2}{((S_1)_i)^2}(S_2)_i
\end{equation}

By substituting the definition of $\widehat{SK}_i$ from Equation~\ref{originalSK} into Equation~\ref{normedSum2}, we may re-express $\overline{S_2}$ as:

\begin{equation}\label{normedSimplifiedSum2}
    \overline{S_2} = \frac{n(n-1)}{n+1}\sum^N_i\left[\widehat{SK}_i\right]  + nN
\end{equation}

We may then construct a multi-receiver spectral kurtosis estimator analogous to that in Equation~\ref{eq:sk},
using the normalized cumulants $\overline{S_1}$ and $\overline{S_2}$ and an effective number of samples $M = nN$:

\begin{equation}\label{avgSK}
\begin{split}
\widehat{SK} 
&= \frac{nN+1}{nN-1} \left ( nN\frac{\overline{S_2}}{(\overline{S_1})^2} - 1 \right) \\
&= \frac{nN+1}{nN-1} \left ( \frac{\overline{S_2}}{nN} - 1 \right) \\
&= \frac{(nN+1)(n-1)}{N(nN-1)(n+1)}\sum^N_i \widehat{SK}_i \\
&= \frac{(nN+1)(n-1)}{(nN-1)(n+1)} Avg(\widehat{SK}_i)
\end{split}
\end{equation}

Equation~\ref{avgSK} illustrates that computing a spectral kurtosis estimate in this manner,
by directly combining samples across the array,
is equivalent to averaging spectral kurtosis estimates generated from the separate receivers.
It should be noted that this is only the case if the individual receivers' data is correctly normalized.
The constant factor in front of Equation~\ref{avgSK} motivates a convenient re-scaling of the multi-receiver spectral kurtosis estimator:

\begin{equation}\label{final}
\begin{split}
\widetilde{SK} 
&= \frac{(nN-1)(n+1)}{(nN+1)(n-1)}\widehat{SK} \\
& = \frac{n+1}{n-1} \left ( \frac{\overline{S_2}}{nN} - 1 \right) \\
& = \frac{1}{N}\sum^N_i \widehat{SK}_i  = Avg(\widehat{SK}_i)
\end{split}
\end{equation}

The statistical properties of this estimator may be directly recovered from those of the constituent $\widehat{SK}_i$; particularly, the expected mean and variance are:
\begin{gather}
 \textrm{mean}(\widetilde{SK}) = 1\\
 \textrm{variance}(\widetilde{SK}) \approx \frac{4}{nN} + \mathcal{O}(\frac{1}{n^2N})
\end{gather}

\subsubsection{Statistical Biases}

The statistical properties described in \S\ref{sec:theory} and \S\ref{sec:MultiFormulation} are precisely correct in the case of continuous sampling of the underlying Gaussian distributions; this requirement is not formally met when implementing the $\widetilde{SK}$ estimator in digital hardware. In correlator systems which use discretized representations of the data, the saturation of the representable range by extreme values and the quantization effects of numerical representations may affect the statistical properties of $\widetilde{SK}$; we discuss this effect in Appendix~\ref{appendix:AppendixB}.

\subsubsection{Computational Cost of the Simplified Multi-Receiver Algorithm}
\label{sec:cmput}

The multi-receiver formulation described above has reduced computational requirements relative to the corresponding individual computations. The minimum number of operations required to calculate a spectral kurtosis estimate for $n$ time-samples and $N$ antennas with and without the combination of receivers are as follows:

Both implementations require several core computations:

\begin{table}[H]
\centering
\begin{tabular}{lcr}
Step & Formula & Operations\\
\hline
Calculating Power & re($x_{ij}$)*re($x_{ij}$) + im($x_{ij}$)*im($x_{ij}$) = $|x_{ij}|^2$ & $3nN$\\
Summing Power in Time& $|x_{i0}|^2 + |x_{i1}|^2 + ... + |x_{in}|^2 = (S_1)_i $ & $(n-1)N$\\
Summing Square Power in Time& ($|x_{i0}|^2)^2 + (|x_{i1}|^2)^2 + ... + (|x_{in}|^2)^2 = (S_2)_i $ & $(3n-1)N$\\
\hline
\textbf{Sub-total} & & \textbf{(7n - 2)N}
\end{tabular}
\end{table}

The additional computations required to produce spectral kurtosis estimates in both cases are as follows:

\begin{table}[H]
\centering
\begin{tabular}{lcr}
\multicolumn{3}{c}{\textbf{Independent $\widehat{SK}$ Estimates}}\\
\hline
Step & Formula & Operations\\
\hline
Previous Sub-Total & & (7n - 2)N\\ 
Calculating $\widehat{SK}_i$ Estimates & $\frac{n+1}{n-1} \left ( \frac{n(S_2)_i}{((S_1)_i)^2} - 1 \right)  = \widehat{SK}_i $ & $8N$\\
Calculating Average $\widehat{SK}_i$  & $ \frac{1}{N}\sum^N_i \widehat{SK}_i =  Avg(\widehat{SK}_i)$ & $N$\\
\hline
\textbf{Total} & & \textbf{(7n + 7)N}\\
\multicolumn{3}{c}{\textbf{Multi-Receiver Formulation}}  \\
\hline
Step & Formula & Operations\\
\hline
Previous Sub-Total & & (7n - 2)N\\ 
Finding Mean Power & $\frac{(S_1)_i}{n} = \mu_i $ & $N$\\
Applying Normalization to Square Power & $\frac{(S_2)_i}{\mu_i^2}  = \overline{(S_2)_i} $ & $2N$\\
Summing of Square Power Across Receivers& $(\overline{S_2})_0 + (\overline{S_2})_1 +...+(\overline{S_2})_N  = \overline{S_2}$ & $N-1$\\
Calculating Multi-Receiver $\widetilde{SK}$ Estimate & $\frac{n+1}{n-1} \left ( \frac{\overline{S_2}}{nN} - 1 \right) = \widetilde{SK}$ & $7$\\
\hline
\textbf{Total} & & \textbf{(7n + 2)N + 6}
\end{tabular}
\end{table}

Although in both cases the leading term is of the form $7nN$, the alternative formulation offers an $\mathcal{O}(N)$ reduction in the number of operations required. For any value of $N$ greater than one, the simplified multi-receiver algorithm requires fewer operations to compute at an identical cadence.

\section{Application to CHIME}
\label{sec:application}

\subsection{Implementation}
\label{ssec:imp}

The CHIME X-engine (\citeauthor{Denman:inprep}, in prep.) consists of 256 correlator nodes, each hosting two dual-chip AMD S9300x2 GPUs. Each of the four GPU chips processes time-ordered data for all of CHIME's 2048 receivers over one $\approx 390$\,kHz frequency channel, and is therefore able to compute $\widetilde{SK}$ estimates for a single sub-band.
The Pathfinder's X-engine has a similar structure; its 16 compute nodes each receive 64 frequency channels from all 256 receivers, and subsequently utilize the same GPU kernels and acquisition software as CHIME to compute $\widetilde{SK}$ estimates.

Two GPU processing kernels were written in Open Computing Language (OpenCL\footnote{\url{https://www.khronos.org/opencl/}}) to compute real-time $\widetilde{SK}$ estimates at a variety of integration lengths. The kernels were then compiled into Heterogeneous System Architecture Code Objects (HSACOs\footnote{\url{https://rocm.github.io/}}) which are executed by CHIME's real-time data processing software, \texttt{kotekan} (\citeauthor{Renard:inprep}, in prep.).
The first kernel computes, accumulates, and normalizes the power and square-power estimates from each independent receiver. The second kernel sums the output from the previous kernel across the array before computing a single $\widetilde{SK}$ estimate for the current time interval. 

This model allows for significant flexibility in the duration of time-integration and total number of receivers in the array, and is highly efficient. In the current implementation, full-array $\widetilde{SK}$ estimates are produced for each of CHIME's 1024 frequency channels after accumulations of 256 time-samples (a cadence of $0.655$\,ms). $\widetilde{SK}$ computations currently require $\approx 0.7\%$ of the available computational power of the CHIME GPU X-engine, and may therefore run in parallel with the primary correlation systems with minimal impact.

\subsection{Results}
\label{ssec:results}

In a June 2018 engineering run,
one hour of simultaneous $\widetilde{SK}$ estimates from both CHIME and the Pathfinder were recorded,
with the values of $\widetilde{SK}$ output at a cadence of $0.655$\,ms. Out of a total 1024 frequency channels across the $400$-$800$\,MHz band, 892 and 272 were recorded to completion on CHIME and the Pathfinder respectively. This data provides a point of comparison between observations and the simulations described in Appendix~\ref{appendix:AppendixB} as well as verifying consistency between the two instruments.

\begin{figure}[H]
\epsscale{.7}
\plotone{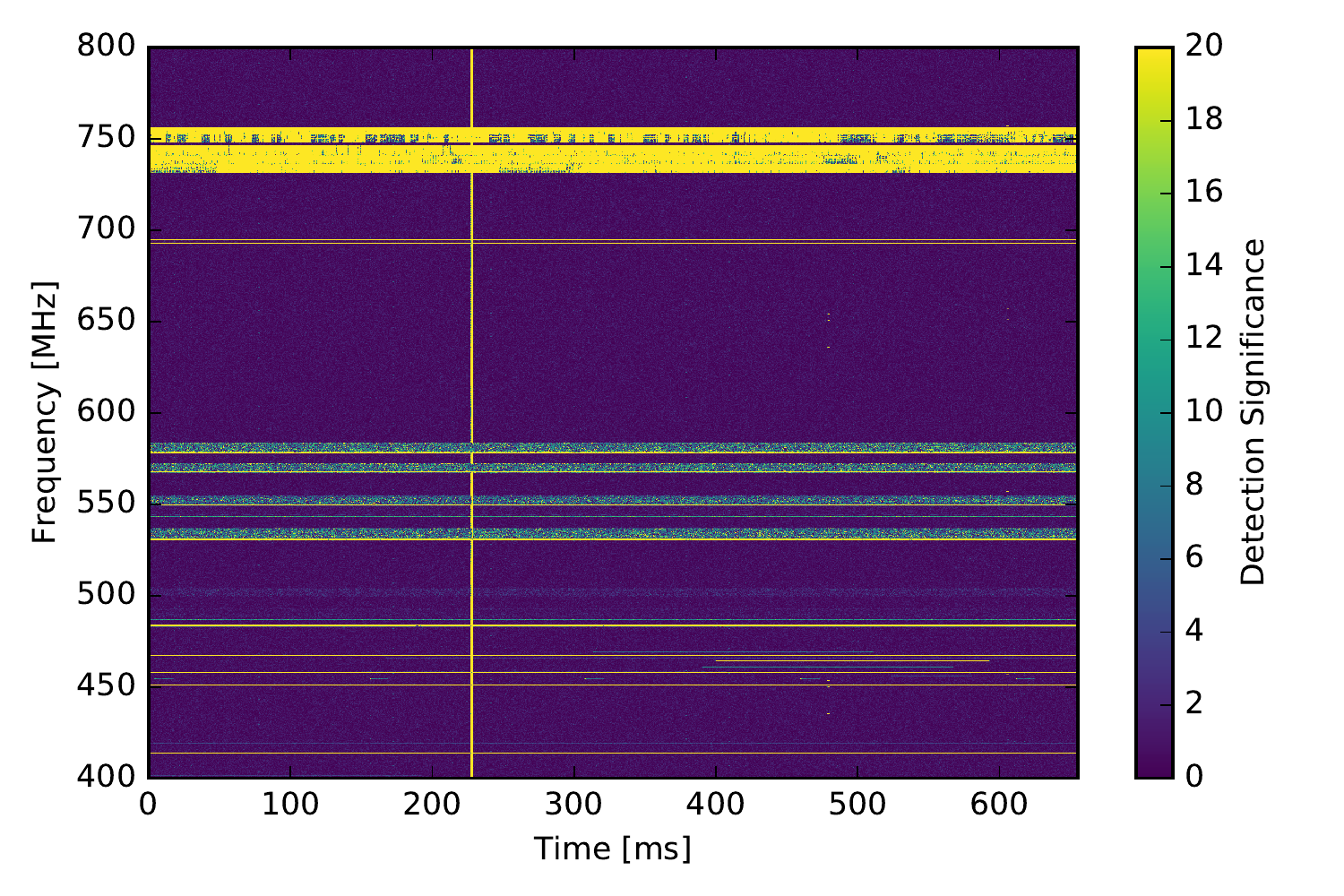}
\caption{\label{fig:ChimePlot} A $\sim 650$\,ms segment of detection confidence values from 892 out of 1024 of CHIME's frequency channels. The broadband pulse (at $\sim 220$\,ms) illustrates the $\widetilde{SK}$ estimator's ability to detect both fixed-frequency and broadband interference; the detection of such pulses highlights the importance of high cadence RFI detection.}
\end{figure}

We define `detection significance' as the difference between the measured value of $\widetilde{SK}$ for a given set of time-samples and its simulated expectation, expressed in units of the standard deviation of $\widetilde{SK}$ when applied to RFI-free data. 
For example, a threshold of `$a$ standard deviations' in detection significance would correspond to removing data outside of the range $\textrm{mean}(\widetilde{SK})\pm a\sqrt{\textrm{var}(\widetilde{SK})}$.
The adoption of a symmetric interval in detection significance as the metric by which RFI is indicated
implicitly relies on the large-$nN$ nature of this implementation;
thresholds based on the predicted false-alarm probability offer an alternative
for smaller-$nN$ cases in which the skewness of the $\widetilde{SK}$ is significant
(see \citet{2016JAI.....541009N} for an example).
Figure~\ref{fig:ChimePlot} depicts a short segment ($\sim650$ms) of `detection significance' values, computed at a cadence of $0.655$ms, for all available frequency channels.
Both fixed-frequency broadcasting signals and a broadband interference pulse ($< 0.655$ms in duration) are detected in the segment. 

\subsubsection{Comparison with Simulated Data}

\begin{figure}[H]
\epsscale{.65} 
\plotone{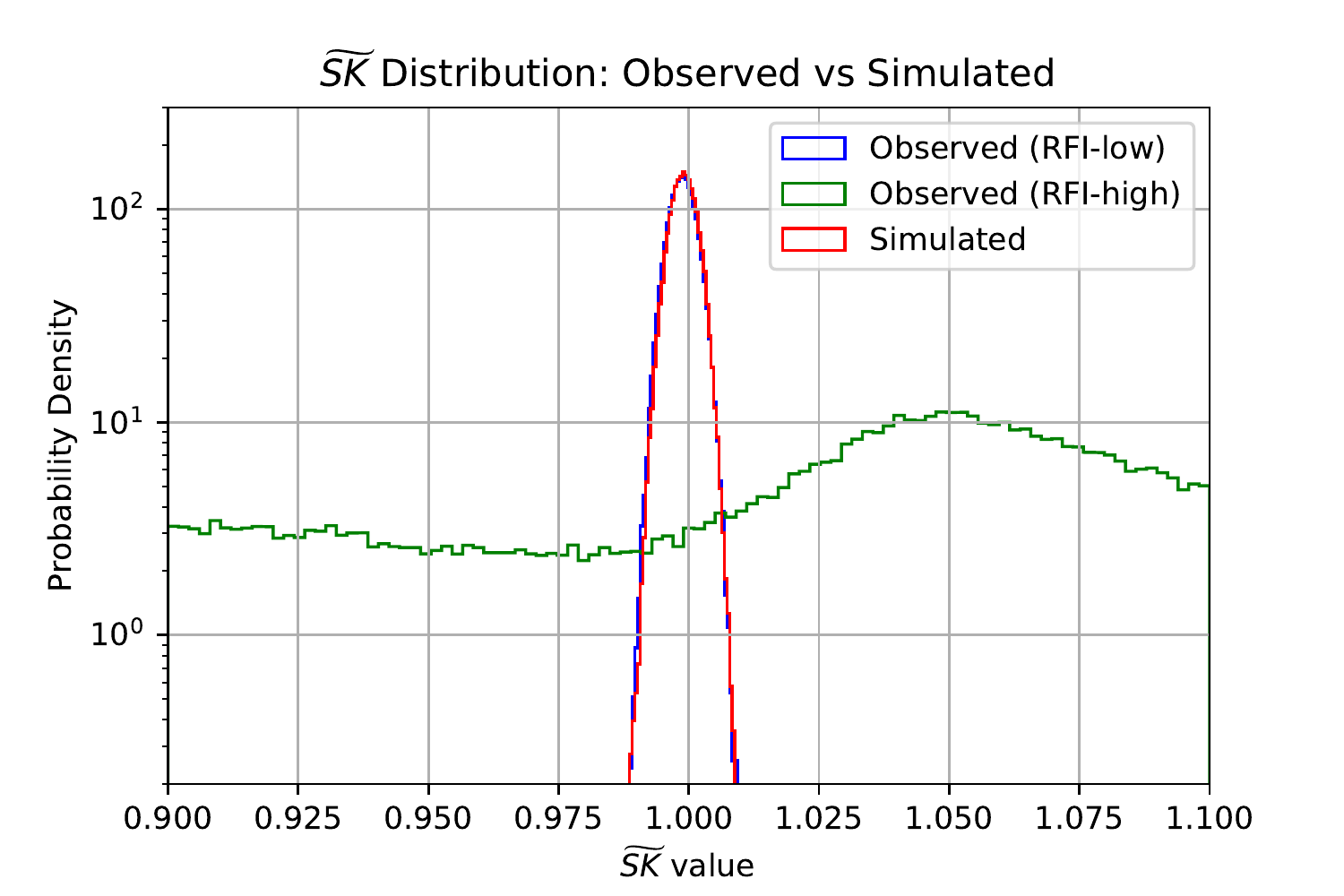}
\caption{\label{fig:skdists} A comparison of the observed $\widetilde{SK}$ distribution of an RFI-free simulation to that of both a highly-contaminated `RFI-high' frequency channel and a relatively clean `RFI-low' channel. The latter matches the simulated distribution's statistics quite closely, while the former has completely inconsistent behaviour. Both the simulated and observed $\widetilde{SK}$ estimates were computed after accumulations of 256 time samples. The simulated and 'RFI-low' distributions have means which are slightly offset from unity as a result of digitization effects, as further discussed in Appendix~\ref{appendix:AppendixB}.}
\end{figure}

\begin{figure}[H]
\epsscale{.7}
\plotone{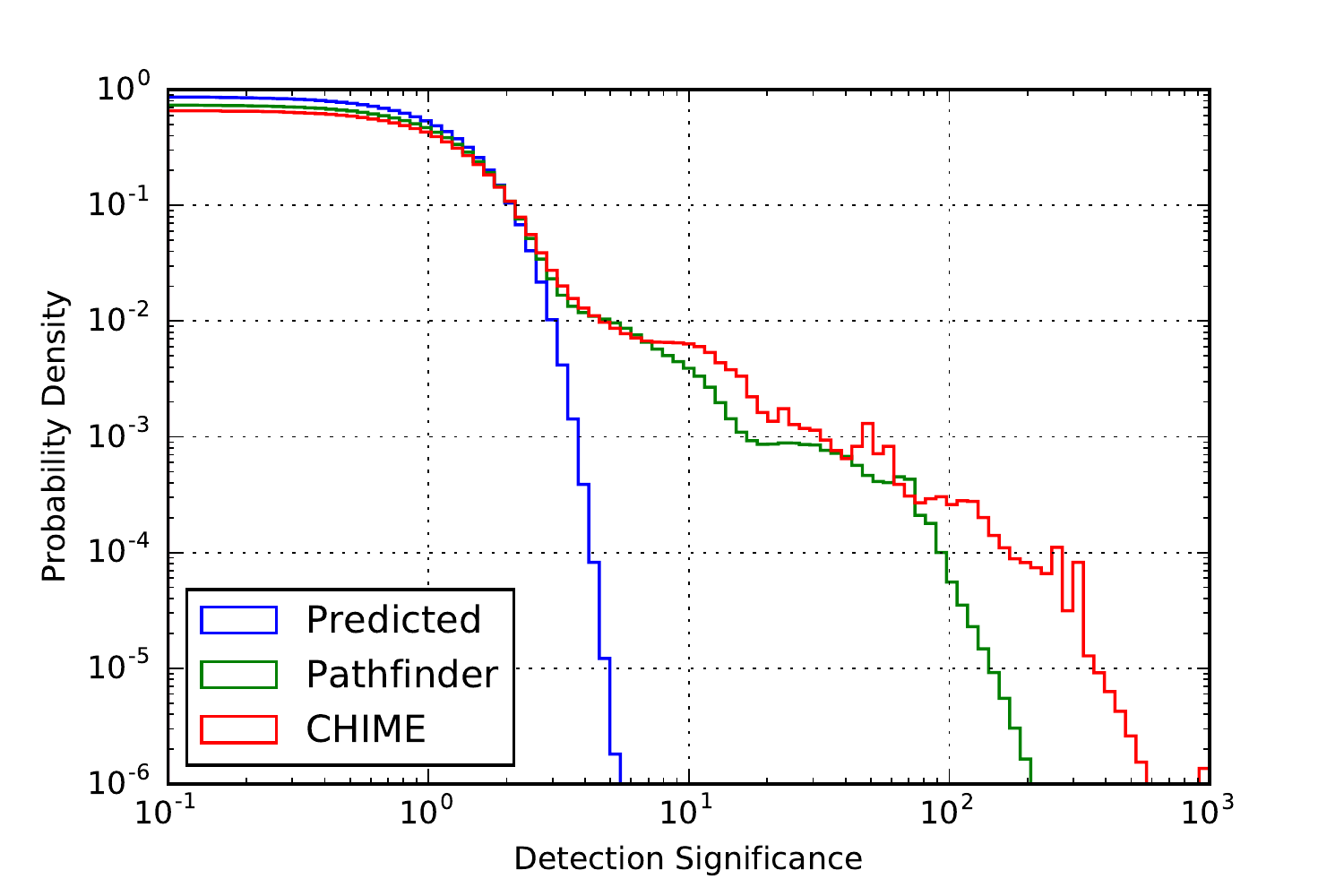}
\caption{\label{fig:confidence}
		The distribution of $\widetilde{SK}$ detection significance 
		in sample data from both CHIME and the Pathfinder,
        as compared to a prediction from simulated RFI-free data.
		The excesses at high significance indicate the presence of 
		RFI in both instruments' data,
		with the rightward shift in the CHIME data
		illustrating the improved interference-to-noise ratio 
		obtained by the larger number of receivers.}
\end{figure}

Data from both CHIME and the Pathfinder indicate that frequencies which are believed to be relatively free of RFI adhere closely to simulations; Figure~\ref{fig:skdists} compares the simulated, RFI-free distribution to CHIME observations of frequency channels with different levels of persistent RFI contamination. Frequencies with low degrees of RFI contamination reproduce the expected $\widetilde{SK}$ distribution while RFI-saturated channels do not. Figure~\ref{fig:confidence} compares the probability distributions of simulated and observed RFI detection significance values (as defined above) in both the CHIME and Pathfinder datasets.  Applying a threshold to this value allows the excision of significantly contaminated time-intervals while also setting a well-constrained false-positive rate for non-contaminated data. When applying a 5-standard-deviation detection threshold, we find that our implementation detects RFI in $\sim$16-18$\%$ of CHIME's band over the course of a day.

\begin{figure}[H]
\epsscale{.7}
\plotone{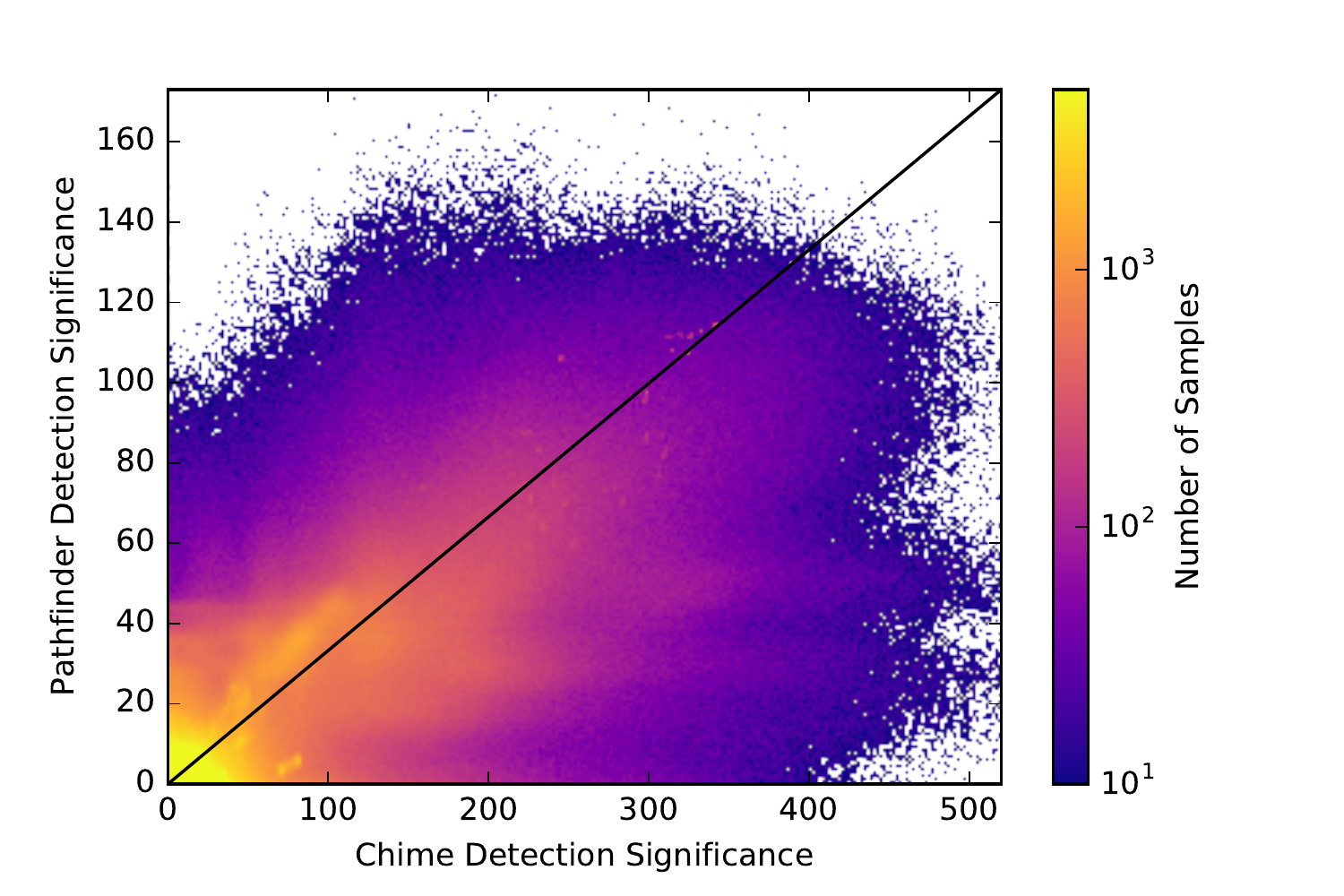}
\caption{\label{fig:heatmap} 
		The simultaneously-measured RFI detection significance
		in synchronous data from both CHIME and the Pathfinder, 
        within the overlapping 216 frequency channels.
		The diagonal line indicates the $\sqrt{N}$ interference-to-noise ratio 
		increase expected from their relative number of receivers.}
\end{figure}

\subsubsection{Inter-Instrument Comparison}

The simultaneous detection of transient events in both CHIME and the Pathfinder
supports the interpretation of changes in $\widetilde{SK}$ as measuring exogenous signals. Figure~\ref{fig:heatmap} presents a comparison of simultaneous $\widetilde{SK}$ estimates from both instruments across all available frequencies; the two are strongly correlated, with a significance ratio consistent with the sensitivity difference predicted by simulations of the two instruments. We conclude that, for a given frequency and time interval, the two instruments provide consistent measurements of the presence of RFI. Moreover, through examination of RFI transient events, it becomes evident that both instruments are observing nearly identical surroundings.

\section{Conclusion}

We have shown that the excision potential of the $\widetilde{SK}$ estimator may be improved through the combination of signals from multiple receivers within a compact array. Our implementation offers a robust, computationally-efficient method for real-time RFI detection and removal for compact arrays. Data from both CHIME and the Pathfinder agree closely with numerical simulations, confirming that the  
effects of quantization on the $\widetilde{SK}$ estimator in our correlator system are entirely predictable. Particularly, we find that these effects do not substantially alter the integration-length-dependence of the estimator's statistical moments. We conclude that the multi-receiver formulation of the spectral kurtosis estimator provides high-time-resolution, low-computational-cost RFI excision whose sensitivity improves predictably with the size of the compact array. 

\section*{Acknowledgements}

We are very grateful for the warm reception and skillful help we have received from the staff of the Dominion Radio Astrophysical Observatory, which is operated by the National Research Council of Canada.
We acknowledge support and funding from 
the Natural Sciences and Engineering Research Council of Canada, 
the Southern Ontario Smart Computing Innovation Platform,
the Ontario Centres for Excellence,
and
the National Research Council of Canada.
We further would like to thank the CHIME collaboration for their support 
and Thoth Technology for the use of the Algonquin Radio Observatory during prototyping stages.
We would finally like to thank the reviewer for their helpful and insightful comments.

\bibliography{rfi}

\pagebreak

\begin{appendices}

\section{Common-Mode Contributions to $\widetilde{SK}$} \label{appendix:AppendixA}
The statistical properties of the spectral kurtosis estimator $\widehat{SK}$ described in \S\ref{sec:theory}, 
depend on the number of samples which are considered. 
The multi-receiver formulation $\widetilde{SK}$, described in \S\ref{ssec:array}, 
combines multiple receivers' samples to reduce the inherent variance in the estimators; 
however, this relies on the independence of each receiver's signal. 
In the presence of sources which contribute substantially to the antenna temperature,
the various receivers may be dominated by a common-mode signal,
introducing a correlation and reducing the effective number of 
independent samples used in computing the $\widetilde{SK}$ estimator.

\begin{figure}[H]
\epsscale{1}
\plottwo{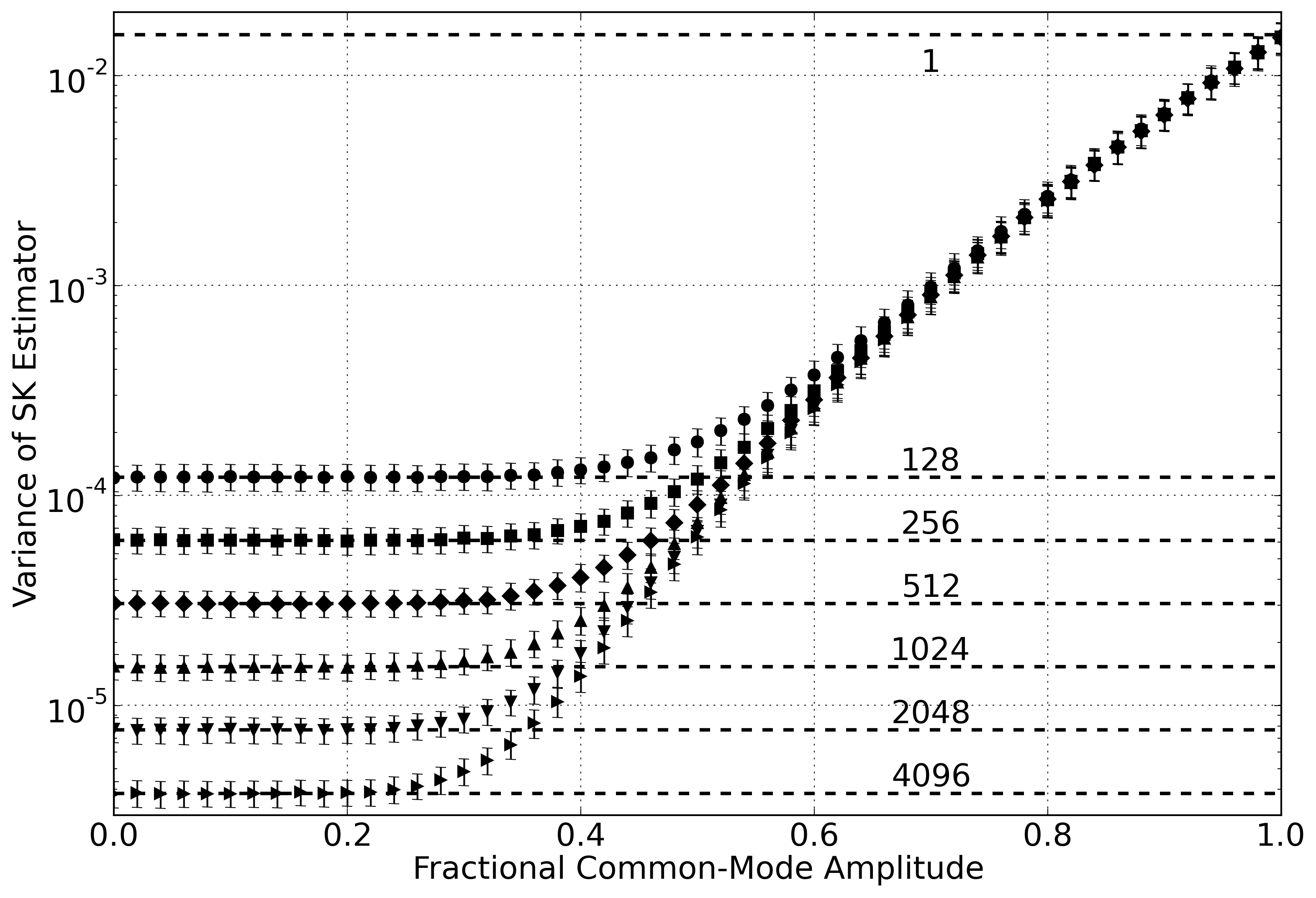}{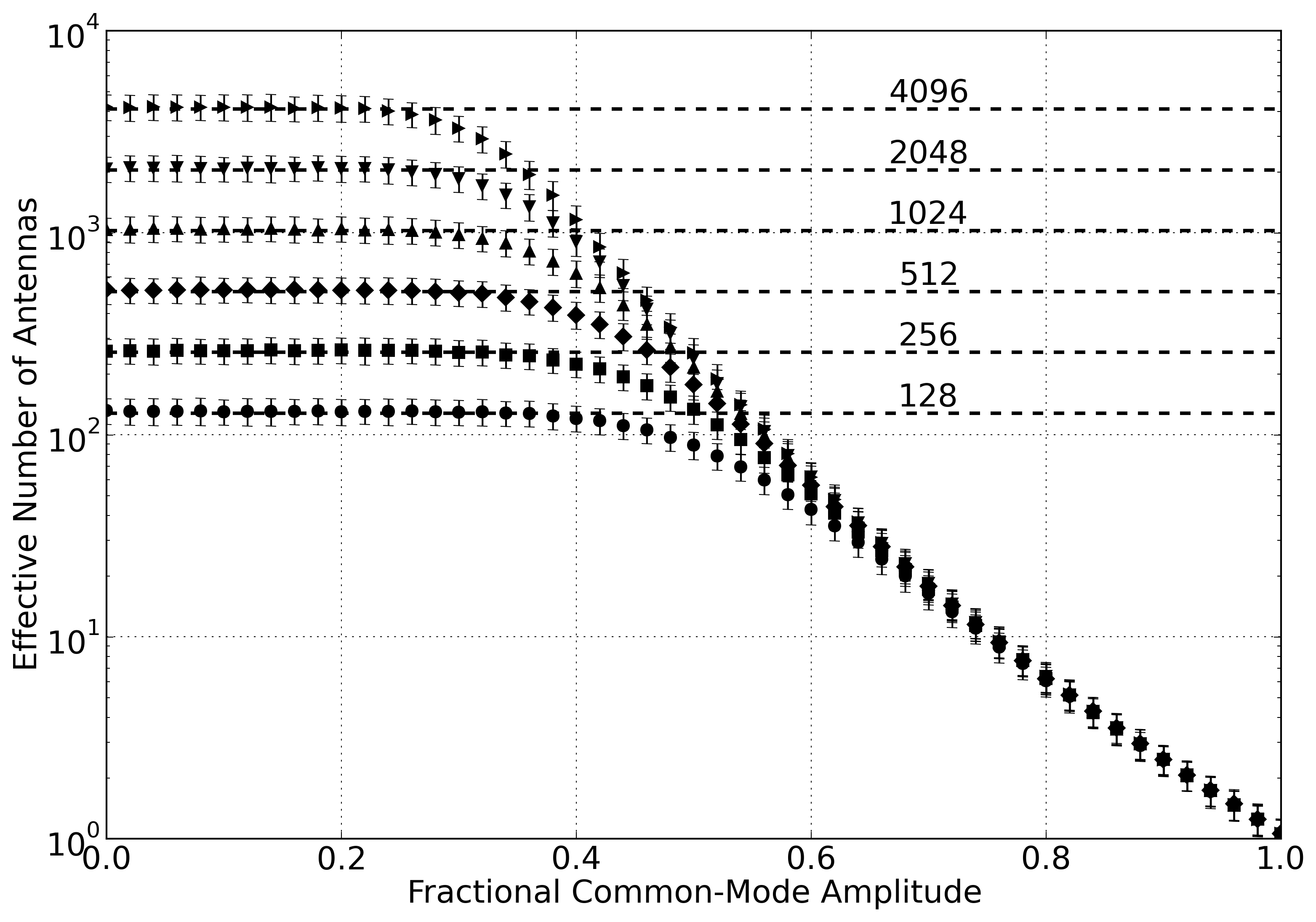}
\caption{\label{fig:common-mode}
Changes in the variance of the multi-receiver $\widetilde{SK}$ estimator in the presence of a common-mode signal.
At left, each point displays the variance in $\widetilde{SK}$ for 1000 datasets, each with $n=256$ time-samples and a number of receivers $N$ ranging from 128 to 4096. 
Dashed lines at top and bottom show the predicted values of $\textrm{var}(\widetilde{SK}) \approx \frac{4}{nN}$ 
for $nN=n=256$ and each of the simulated values of $N$.
At right, the variance is re-expressed as the effective number of receivers $N = \frac{4}{n\cdot\textrm{var}(\widetilde{SK})}$, which smoothly declines from the full complement to an effectively-single-receiver case when the common-mode signal dominates the system.
}
\end{figure}

Figure \ref{fig:common-mode} shows the results of a numerical simulation in which a number of independent Gaussian datasets, representing the noise-dominated receiver timestreams, are augmented by a common-mode Gaussian signal representing the `source'. As the common-mode signal's amplitude is increased, the variance in $\widetilde{SK}$ increases, eventually reaching the value which would be predicted for a single-receiver timestream with identical properties. Notably, larger-$N$ arrays see this effect at lower levels of common-mode signal, placing a sensitivity-dependent upper limit on the size of array for which the multi-receiver formalism will show practical benefits.

In the case of CHIME, the instantaneous correlation between receivers is rarely above a few percent - the exception occurs at solar transit, where correlations of order unity are observed at some frequencies. As data is generally of low quality during solar transit, and is discarded, this effect is considered negligible for our purposes.

\section{Digitization Effects on $\widetilde{SK}$} \label{appendix:AppendixB}

The quantization of data within a telescope's digital processing systems may have significant effects on the properties of spectral kurtosis estimators as applied to the resultant data.
\citet{7833535} examine this effect and address it by constructing a Gamma distribution with empirically-derived shape and scale parameters, 
and then using the properties of this function to set the appropriate shape factor for their generalized spectral kurtosis estimator.
In our implementation, 
given the large number of samples accumulated 
and the well-constrained properties of data within the CHIME digital signal processing system,
we have opted to generate excision thresholds 
based directly on numerical simulations of the CHIME correlator's data characteristics.

\begin{figure}[H]
\epsscale{1}
\plottwo{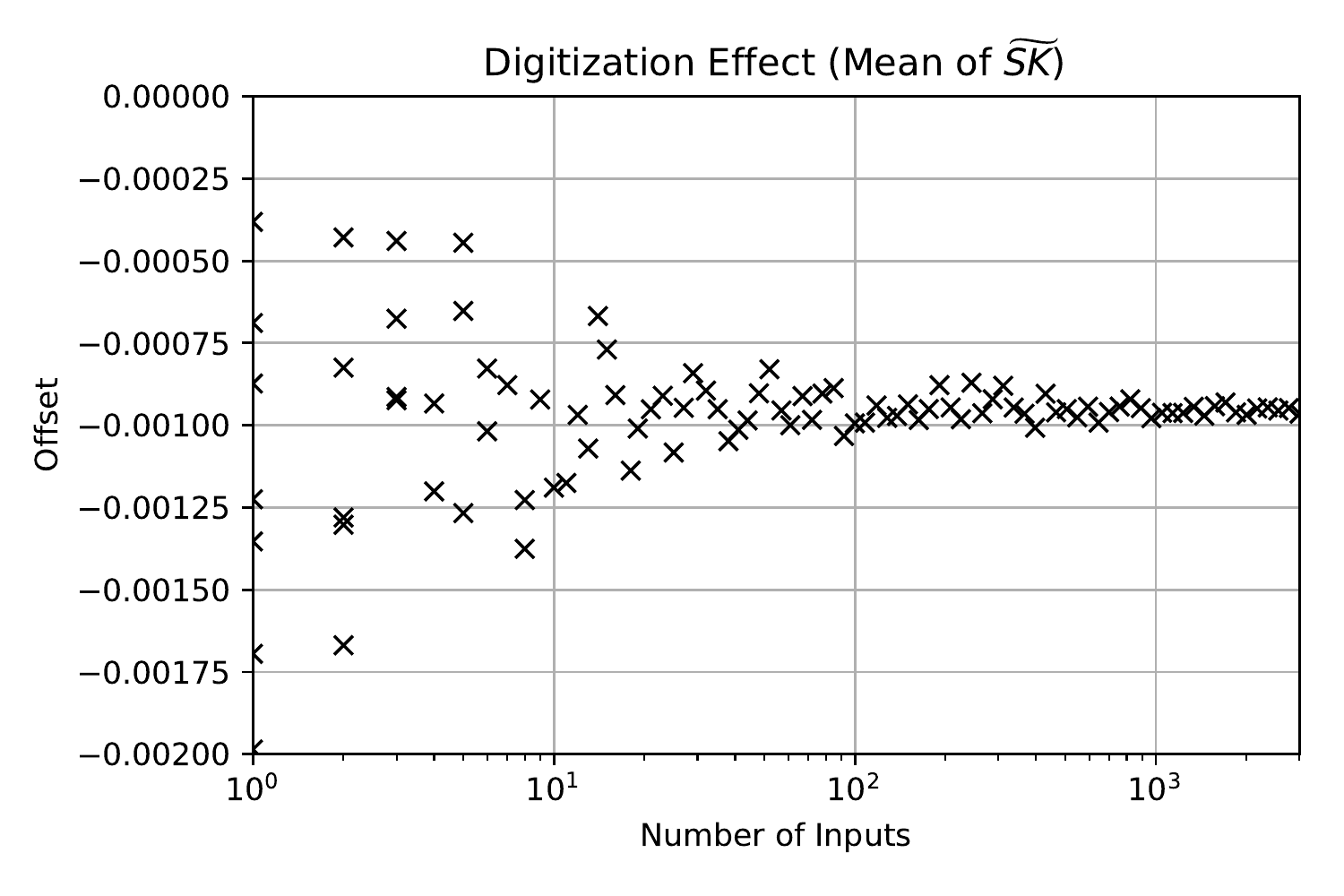}{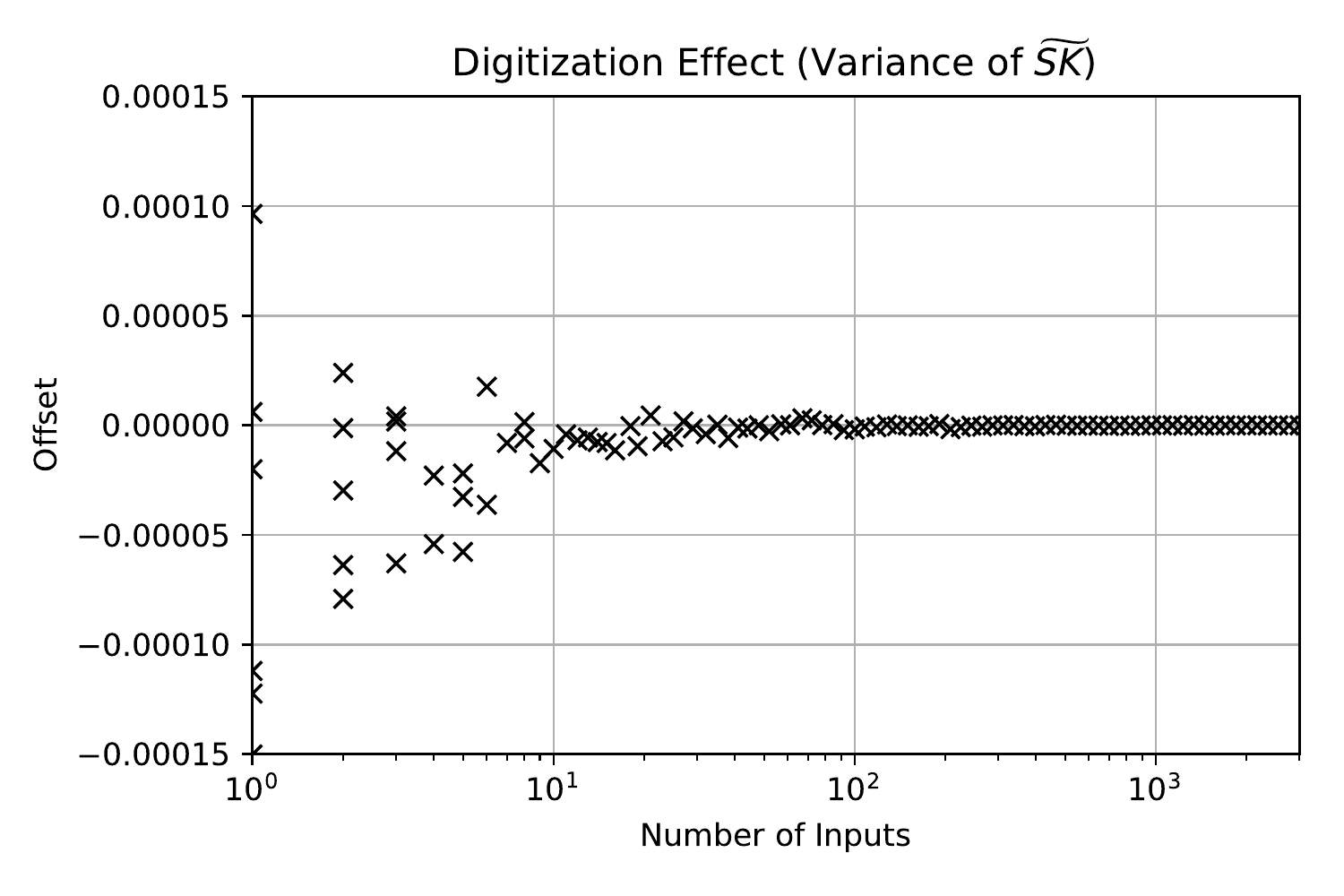}
\caption{\label{fig:SimulatedStats} Differences between the statistical moments of `digitized' (heavily discretized) and non-digitized $\widetilde{SK}$ estimates for a variety of effective array sizes. For large numbers of inputs, the mean (left) converges to a value which is offset from the expected result by a constant, while the variance (right) converges to the same value in both digitized and non-digitized simulations.}
\end{figure}

To simulate the digitization-related truncation and rounding effects present in the CHIME correlator, a set of sample data was generated which mimicked the properties of CHIME's intermediate data products -- 4+4-bit complex Gaussian data (bounded between $-7$ and $7$) with an RMS of $\approx1.52$ least-significant bits in each component. Following \S\ref{sec:MultiFormulation}, $\widetilde{SK}$ estimates were computed for a variety of integration lengths and array sizes. The statistics of those estimates were averaged over multiple iterations and compared to analogous non-`digitized' (floating-point) simulations, the results of which are presented in Figure~\ref{fig:SimulatedStats}. 

In the case of CHIME-like discretization, digitization effects on the variance (and higher moments) of the $\widetilde{SK}$ estimator vanish for large numbers of inputs. 
The mean value of the $\widetilde{SK}$ estimator, however, converges to a value slightly offset from unity.
This offset is added directly to the value of the RFI excision threshold;
as the higher moments are unchanged for the large number of inputs considered,
this restores behaviour identical to the threshold-determination methods described in \S\ref{ssec:results}.

\end{appendices}
\end{document}